\documentclass[pra,onecolumn,amsmath,amssymb]{revtex4}
\usepackage{amsfonts}
\usepackage{graphicx}
\usepackage{dcolumn}
\usepackage{float}
\usepackage{bm}

\newcommand{\ind}[1]{\textrm{#1}}
\newcommand{\av}[1]{\langle {#1} \rangle}
\newcommand{\ket}[1]{| {#1} \rangle}
\renewcommand{\O}{ {\cal{O}} }
\newcommand{\com}[1]{}
\newcommand{\J}{{\cal J}}
\newcommand{\B}{{\cal B}}
\newcommand{\D}{{\cal D}}
\renewcommand{\S}{{\cal S}}

\begin{document}

\title{Cold Atom Qubits}

\author{Dmitry Solenov\footnote{E-mail: solenov@lanl.gov (the author to whom the correspondence should be addressed)} and Dmitry Mozyrsky\footnote{E-mail: mozyrsky@lanl.gov}}

\affiliation{Theoretical Division (T-4), Los Alamos National
Laboratory, Los Alamos, NM 87545, USA}

\date{\today}

\begin{abstract}

We discuss a laser-trapped cold-atom superfluid qubit system. Each
qubit is proposed as a macroscopic two-state system based on a set
of Bose-Einstein condensate (BEC) currents circulating in a ring,
cut with a Josephson barrier. We review the effective low energy
description of a single BEC ring. In particular, it is
demonstrated that such system has a set of metastable current
states which, for certain range of parameters, form an effective
two-state system, or a qubit. We show how this qubit can be
initialized and manipulated with currently available
laser-trapping techniques. We also discuss mechanisms of coupling
several such ring qubits as well as measuring individual
qubit-ring systems.
\end{abstract}


\maketitle

\section{Introduction}

Since the first successful experiments on Bose-Einstein
condensation of alkali gases, optical cooling and trapping have
become a standard experimental tool to study quantum degenerate
gasses at ultra-low temperatures. The laser trapping of neutral
atoms has been rapidly developing over the past decade.
State-of-the-art traps are capable of forming precise time-varying
potentials with micron-size resolution \cite{Boshier}. High
densities ($\sim 10^{10-13}$ cm$^{-3}$) and wide range of
interactions have been achieved at nK temperatures
\cite{Rb85exp,Ketterle-lowT}. Not surprisingly, cold atom systems
have become a perfect tool to test fundamental physics from basic
quantum mechanical principles and quantum information to complex
strongly correlated quantum states of matter
\cite{Lehur,Kalas1,Kalas2,Eddy,SM-BF1,SM-BF2}.

Trapping of neutral atoms is based almost entirely on resonant
scattering of light. It utilizes the internal energy structure of
an atom to the control radiation pressure of the trapping
(cooling) laser beam. An accurately configured laser field creates
an effective (attractive or repulsive) potential profile for each
atom, see
\cite{Ketterle-lowT,Chu,Cohen-Tannoudji,Phillips,LeggettBEC} for
further details. Each atom is indistinguishable and has either
integer or half-integer spin depending on its nuclear and
electronic content. Hence, at the temperatures of typical
cold-atom experiment, atoms obey Bose or Fermi statistics, e.g.
typical ``bosons" are $^{87}$Rb, $^{85}$Rb, $^7$Li, $^{133}$Cs,
etc.

Quantum information has been introduced in a cold-atom system on
various levels \cite{Phillips2,Lundblad}. It is natural, for
example, to define a qubit (two state system) via two internal
states of an atom. In this case, each atom records a single bit of
quantum information. This approach, however, requires each atom to
be addressed separately. A similar problem appears when the qubit
is introduced via a set of spatially localized states (e.g. in
adjacent wells of an optical lattice potential) of an atom or a
dilute Bose-Einstein Condensate (BEC). The complication is due to
the fact that the number of atoms, $N$, in a BEC (cold-atom)
experiment fluctuates significantly, about 10\% (or at least as
$\sqrt{N}$), from run to run. Therefore any qubit system dependent
on the number of atoms becomes problematic.

Another way to introduce a two-state system is via a collective
phenomena. In this case the qubit has to be formed by a pair of
distinct macroscopic states that are sufficiently far away from
other states of the multiparticle system and, at the same time,
have the energy difference between these lowest states small
enough to allow measurable dynamics \cite{LeggettFrance}. Already
in the early stages, the experiments on BEC of trapped atoms were
focused on the dynamics in a double-well(dot) trapping potential
\cite{AndrewsKetterle}. However, despite the appealing similarity
with a single-particle two-state system, the BEC system of such
geometry presents no easy way to achieve superposition of
macroscopic states with measurable two-state dynamics. Consider,
for example, a typical system of non-interacting (weakly
interacting) boson atoms in a double-well confinement potential.
At low temperatures (typically $\sim 100$ nK) the atomic gas forms
a BEC state, which is a many-particle product state $\prod_i
[A\chi_L(\mathbf{r}_i)+B\chi_R(\mathbf{r}_i)]$, where
$\chi_{L/R}(\mathbf{r})$ is a single particle state localized in
the left/right well. This wave function is clearly not suitable to
define a microscopic two state system---it describes a collection
of non-interacting indistinguishable microscopic (single particle)
two state systems that have been discussed earlier. A similar
system with repulsive scattering favors a Mott state of type
$\mathcal{P}[\prod_{i=1}^{N/2}\chi_L(\mathbf{r}_i)]
[\prod_{i=N/2+1}^N\chi_R(\mathbf{r}_i)]$, where $\mathcal{P}$
denotes permutation of particle indexes. A schrodinger-cat state
$A\prod_i\chi_L(\mathbf{r}_i)+B\prod_i\chi_R(\mathbf{r}_i)$
appears only in the limit of sufficiently strong attractive
interaction. Such a two-state system, however, is difficult to
handle. The transitions between the left and right macroscopic
states are greatly suppressed. The tunneling exponent is
proportional to the number of particles. Coherent evolution
appears on reasonable time scales only in the limit of vanishingly
small barrier.

The outlined difficulties can be overcome in a two-state system
defined by macroscopic states that span the same spatial region,
e.g. superfluid (BEC) current states. The simplest geometry
featuring such persistent current states is a ring. It is inspired
by the celebrated superconducting flux qubit
\cite{LeggettFrance,squid,amb}---a superconductor ring with a
Josephson junction. In a superconductor ring system the qubit is
encoded by the magnetic flux states which are due to the
supercurrent of charged particles (Cooper pairs of charge 2e). Low
energy dynamics of these states is due to an interplay of three
energies: the magnetic inductive energy of the ring, the electric
field charging energy, and the energy associated with the
Josephson junction---a cut in the superconducting ring. Neutral
atomic BEC confined in a ring featuring weak link (Josephson
junction) does not have the first two of these energies.
Nevertheless, it turns out that such BEC system can still mimic
the behavior of superconducting Josephson ring device in certain
cases \cite{SolenovMozyrskyRING}.

In the following we address the essential physics of the BEC
Josephson ring system. We derive an effective low energy
description (Sec.~\ref{sec:BECJ}) and show that it is suitable to
introduce a macroscopic two-state system, or qubit
(Sec.~\ref{sec:BECQ}). We discuss possible approaches to single
and multi-qubit dynamics (Sec.~\ref{sec:MQ}) and outline
measurement procedures (Sec.~\ref{sec:Measure}). We do not discuss
the advantages of BEC-Josephson qubit systems over more
traditional qubit designs. The goal of this investigation is to
introduce an approach to quantum information based on trapped BEC
persistent-current states. While we give some analysis of a
BEC-based qubit and discuss basic quantum operations, we do not
perform a comprehensive study of single- and multi-qubit gates.
Rather, we outline basic principles and state challenges that will
have to be addressed to fully understand the potentials of this
system.

\section{BEC-Josephson system}\label{sec:BECJ}

In this section we present a detailed review of a single
persistent-current BEC-Josephson device. The system is inspired by
celebrated superconducting Josephson ring---the simplest geometry
of a superconductor flux qubit \cite{LeggettFrance}. The starting
point for our system is a toroidal BEC \cite{tor1,tor2,tor3}---a
BEC of neutral atoms confined in a ring-shaped potential. Since
the atomic system is in a Bose-condensed state it maintains phase
coherence along the ring. As the result, the phase flux through
the ring can change only by multiples of $2\pi$, i.e. $2\pi\nu$.
The integer $\nu$ is a winding number. It defines quantized
current states: recall that the superfluid current is proportional
to the gradient of phase along the ring. To have a non-integer
winding number the phase has to jump discontinuously at some point
along the ring (phase slip). This is possible only if the BEC
density at that point vanishes. The energy cost of creating
vanishing density, or a node, comes from kinetic energy and is
usually high. However, if a sufficiently high potential barrier is
introduced, the node can be easily created inside the barrier. As
the result, the phase across the barrier can change only by a
fraction of $2\pi$, causing a small increase in energy, $\sim E_J$
(Josephson energy). The interplay of this energy with the energy
of the BEC in the bulk of the ring creates rich potential
landscape which can be exploited to create and manipulate
persistent current states. In the following we will derive an
effective low-energy action suitable to capture essential physics
of persistent current states. The derivations will largely follow
Ref.~\cite{SolenovMozyrskyRING}.

\begin{figure}
\includegraphics[width=10.0cm]{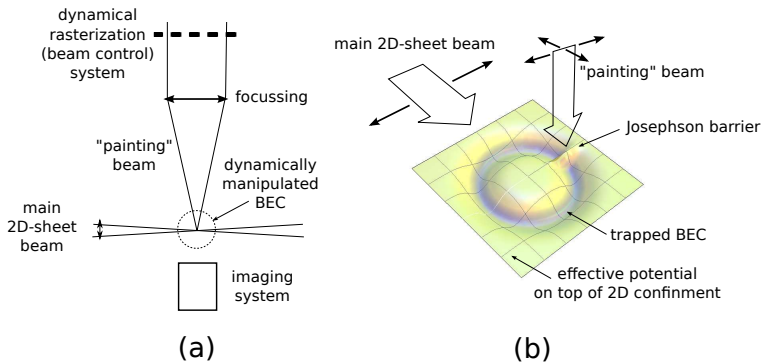}
\caption{(a) Essential components of dynamically manipulated
"painted potential" trap \cite{Boshier}. (b) Ring-shaped BEC
(schematically) formed in the toroidal potential well cut with the
Josephson barrier. The effective (time-averaged) potential profile
is created by the "painting" beam on top of 2D-sheet confinement.
}\label{bec-ring.eps}
\end{figure}

The system of interest is shown in Fig.~\ref{bec-ring.eps}, see
also Ref~\cite{Boshier}. As soon as atoms are cooled down to nK
temperatures the trapping is achieved by two laser breams. With
the appropriate detuning they can be set effectively attractive
(potential is lower in region with higher light intensity). The
main beam is directed horizontally, see
Fig.~\ref{bec-ring.eps}(a), and focused to a few $\mu$m in the BEC
region. It oscillates to the sides with frequency $\sim 100$ Hz.
The atoms move much slower and observe only a time-averaged beam.
A virtually flat 2D-plane potential well is created. The other
``painting" beam scans the 2D-sheet trap from above creating
time-averaged effective attractive or repulsive potential in
addition to the primary 2D trap. The potential profile
corresponding to the BEC-Josephson system is (schematically) shown
in Fig.~\ref{bec-ring.eps}(b). It is a toroidal potential well of
radius $10-100\mu$m (circumference $L\sim 60-600\mu$m) and
thickness of about $1-10\mu$m. The $1\mu$m-size potential barrier
is placed at some point of the toroidal well. The barrier can
rotate along the ring with variable frequency (typically $\sim$
Hz) to stir the condensate.

The trap with parameters outlined above permits several regimes of
BEC dynamics: (a) Complex three-dimensional (3D) flow. If the
thickness of the BEC ring is significant compared to its healing
length, $\xi$ , vortex excitations can form in the bulk of the BEC
ring. This regime has been explored in \cite{Collins}. (b)
Quasi-one-dimensional (1D) flow along a large-circumference ring.
This regime is realized when vortex excitations are squeezed to
higher energies by tight (of the order of few $\xi$) transverse
confinement, while the circumference of the ring is large so that
the kinetic energy of atomic motion (BEC flow) corresponding to
$\nu\sim 1$ is negligible. (c) Quasi-1D small ring limit. In this
case kinetic energy of rotation plays an important role. We will
focus on the quasi-1D regime of BEC dynamics, i.e. (b) and (c).

We start with the microscopic description of a quasi-1D BEC system
and investigate the partition function, $Z$, at low temperatures,
\begin{equation}\label{eq:Z-def}
Z = Tr[e^{H/k_BT}] = \int\D\psi^*\D\psi e^{-\S},
\end{equation}
where $\psi$ is the boson field variable. Our goal is to formulate
the effective action in terms of  the phase difference across the
junction $\phi = \varphi (0) - \varphi (L)$ integrating out other
degrees of freedom. All the parameters are assumed in the units of
a quasi-1D system; we also set $\hbar=1$. The microscopic action
$\S$ is
\begin{equation}\label{eq:S-def}
\S=\S_0 + \S_\J,
\end{equation}
\begin{equation}\label{eq:S0-def}
\S_0=\int_0^\beta d\tau\int_0^Ldx \psi^*(x,\tau)\left[
\partial_\tau -\frac{\nabla^2}{2m} +
\frac{\lambda}{2}\psi^*(x,\tau)\psi(x,\tau)\right]\psi(x,\tau),
\end{equation}
\begin{equation}\label{eq:SJ-def}
\S_\J = \int_0^\beta d\tau\J\left[\psi^*(0,\tau)\psi(L,\tau) +
\psi^*(L,\tau)\psi(0,\tau)\right].
\end{equation}
Here $\beta=1/k_BT$, and the position along the ring, $x$, is
measured from the junction edge. It is convenient to introduce
parametrization
$\psi(x,\tau)=\sqrt{\rho+\delta\rho(x,\tau)}e^{i\varphi(x,\tau)}$,
where $\rho$ is the mean-field BEC density. This density is
constant, $\rho = N/L$, everywhere along the ring except near the
Josephson barrier where it approaches zero. Spatial variations of
density in this region are lumped into the effective Josephson
constant $\J$. With these definitions we obtain the standard
Josephson action
\begin{equation}\label{eq:SJ}
\S_\J = \int_0^\beta d\tau E_J\cos\phi(\tau),
\end{equation}
where $E_J=\J N/{L}$ is Josephson energy. Calculations for the
bulk action $\S_0$ are more involved. We first rewrite $\S_0$ in
terms of fluctuation fields
\begin{equation}\label{eq:S0-expand}
\S_0=\S_{MF} + \int_0^\beta d\tau\int_0^Ldx \left[ i\delta \rho
\dot \phi  + \frac{\rho}{2m}(\nabla\phi)^2 + \frac{\lambda
}{2}\delta \rho ^2 + \frac{1}{8\rho m}(\nabla\delta\rho)^2 + ...
\right].
\end{equation}
The low-energy physics is governed by long-wave fluctuations
\cite{SolenovMozyrskyRING,Buchler}, i.e. $\nabla$-terms and also
$\dot\phi$ are considered small. Another small parameter is the
magnitude of the density fluctuations. Therefore it is appropriate
to keep only the first three terms under the integral sign in
Eq.~(\ref{eq:S0-expand}). After this approximation the density
fluctuations enter in quadratic form, $Z\sim\int\D\delta\rho
\exp[-A\delta\rho^2]$, and can be integrated out. We obtain
\begin{equation}\label{eq:phonon-action}
\S_0=\S_{MF} +  \frac{1}{L}\sum_k\int \frac{d\omega}{2\pi}
\frac{1}{2\lambda}\left[c^2k^2+\omega ^2
\right]\varphi^2(k,\omega),
\end{equation}
where the velocity of sound is $c^2  = \rho\lambda_B/m$. This is
the standard phonon action formulated in terms of phase
fluctuations (density and phase fluctuations are conjugate
variables). We will focus on the tractable limit for which sound
fluctuations along the ring are bounded by the Josephson junction
and, hence, quantized by the boundary condition
$\partial_x\varphi(0,\tau)=\partial_x\varphi(L,\tau)=0$. This is
appropriate as soon as the level spacing $\min(ck) = 2\pi c/L$ in
the excitation energy spectrum is much larger than the Josephson
energy per particle $E_J/N$, i.e. $c\gg\J$. In this limit the
action (\ref{eq:phonon-action}) can be reduced to a local action
by integrating out $\varphi^2(k,\omega)$ assuming that
$\phi(\omega)=\varphi(0,\omega)+\varphi(L,\omega)$. It is
accomplished by introducing the functional $\delta$-function
according to
\begin{equation}\label{eq:Z-delta}
Z = \prod\limits_\omega\int{\mathcal{D}\phi \mathcal{D}\varphi}
e^{ - \S} \delta[\phi(\omega)-\varphi(0,\omega)+\varphi(L,\omega)]
\end{equation}
and using the identity $\delta(a)=(2\pi)^{-1}\int d\Lambda
\exp{(i\Lambda a)}$. The fields $\varphi$ and $z$ can then be
integrated out. Note that in doing so we expand field $\varphi(x,
\tau)$ in $\cos{(\pi n x/L)}$, thus satisfying the boundary
conditions above. After some calculation we notice that the
partition function takes the form $Z=\int \mathcal{D}\phi
\exp{(-\S_{\rm eff})}$, where
\begin{equation}\label{eq:effective-action-full}
\S_{\rm eff} = \int\frac{d\omega}{2\pi} |\phi(\omega)|^2
\frac{c\,\omega}{4\lambda\tanh{\left(\frac{\omega L}{2c}\right)}}
-E_J\int d\tau\cos\phi(\tau).
\end{equation}
In the limit of large ring, $L\to\infty$ (regime (b) as discussed
earlier), we recover the Caldeira-Leggett dissipative action
\cite{LeggettChakravarty,weiss}
\begin{equation}\label{eq:effective-action-CL}
\S_{\rm eff} \to \frac{c}{4\lambda} \int\frac{d\omega}{2\pi}
|\omega| |\phi(\omega)|^2 - E_J\int d\tau\cos\phi(\tau),
\end{equation}
which describes dissipative particle dynamics in a periodic
potential.
\begin{figure}
\includegraphics[width=11.0cm]{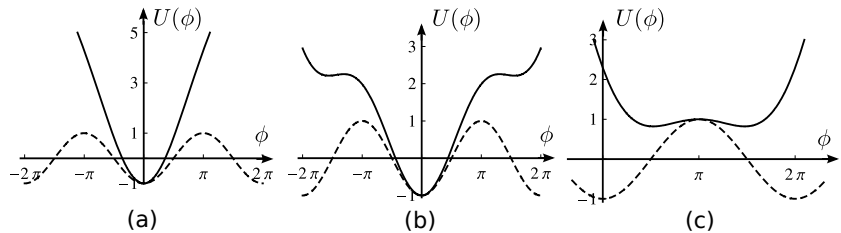}
\caption{Typical effective potential energy profiles (in units of
$E_J$) as functions of the phase across the Josephson barrier. For
comparison, pure Josephson energy contribution is given. (a)
Josephson energy, $E_J$, is small compared to effective the
``inductive" energy (kinetic energy of rotating BEC). (b) $E_J$ is
comparable to the effective ``inductive" energy. (c) Same as (a)
but with rotating Josephson barrier (in rotating frame of
reference).}\label{Uiii.eps}
\end{figure}

We are interested in the ``coherent" limit when $L$ is finite and
$\omega\to 0$ (regime (c) as discussed earlier). In this case we
obtain
\begin{equation}\label{eq:effective-action-coherent}
\S^0_\ind{eff} = \int d\tau \left[\frac{L}{24\lambda}
\dot\phi(\tau)^2 + \frac{\rho}{2mL}\phi(\tau)^2 - E_J
\cos\phi(\tau)\right].
\end{equation}
Note that all the terms in
Eq.~(\ref{eq:effective-action-coherent}) resemble those of the
superconducting flux qubit. In the latter system, however, the
first two terms originate from classical electromagnetism
\cite{LeggettFrance}: the $\dot\phi$-term,
$(\hbar^2C/8e^2)\dot\phi^2$, is due to the electric field of the
capacitor formed by different sides of the Josephson junction (and
the ring); the $\phi^2$-term, $(\hbar^2/8e^2\mathcal{L})\phi^2$,
is the result of the magnetic field energy induced by the
supercurrent. In our case the kinetic term is the consequence of
phonon quantization, while the $\phi^2$-term is due to the kinetic
energy of the collective motion of the particles---superfluid
current.

Finally, we consider the case of a rotating Josephson barrier. At
this point it is already clear that the effect of rotation should
contribute to the $\phi^2$-term of the above action. As we will
see shortly, the rotation of the trap is indeed equivalent to a
flux of external magnetic field used to tune the superconducting
Josephson rings. Consider the rotation \cite{LeggettBEC} of the
form $V(x\cos(\Omega t)+y\sin\Omega t,x\cos(\Omega t)-y\sin\Omega
t,z)$. In the rotating frame of reference $x\to x+vt$, $v=\Omega
L/2\pi$. This change adds the term $\int d\tau\rho v\phi(\tau)$ to
the above action. Introducing $\alpha = m\J L$ we obtain
\begin{equation}\label{eq:effective-action-coherent-rotating}
\S^0_\ind{eff} = E_J\int d\tau \left[\frac{L}{24\lambda E_J}
\dot\phi(\tau)^2 + \left\{\frac{(\phi(\tau)-\phi_0)^2}{2\alpha} -
\cos\phi(\tau)\right\}\right],
\end{equation}
where $\phi_0 = mL^2\Omega/2\pi$. The parameter $\alpha$, together
with $\phi_0$, determines the shape of the effective energy, i.e.
the last two terms under the integral sign. Three situations are
possible, see Fig.~\ref{Uiii.eps}: (a) When the tunneling through
the Josephson barrier is negligible ($\alpha\to 0$) the atomic
system condenses into the rotating BEC state defined by $\phi_0$.
(b) When the tunneling becomes significant the stationary system
($\Omega=0$) develops a set of metastable current-carrying states
(local minima of the effective potential). The metastable states
appear when both the first and the second derivative of the
effective potential with respect to $\phi$ vanish. The
corresponding set of equations is
\begin{equation}\label{eq:metastable-alpha}
\tan\phi_{s,i}=\phi_{s,i},\quad\quad\quad \alpha_i =
-1/\cos\phi_{s,i}.
\end{equation}
The first pair of metastable states appear when
$\alpha>\alpha_1\approx 4.6$ ($\phi_{s,1}\approx\pm 1.43\pi$), the
second when $\alpha>\alpha_2\approx 10.95$, and so on. (c)
Finally, when rotation $\phi_0\sim\pi$ (winding number $1/2$) and
$\alpha>1$ a double-well potential forms, see Fig.~\ref{Uiii.eps}.
In this review we are interested in the latter case. The analysis
of the previous two situations can be found in
\cite{SolenovMozyrskyRING}.

\section{BEC current-based two-sates system: BEC
qubit}\label{sec:BECQ}

\subsection{Formulation of a two-state system}
\begin{figure}
\includegraphics[width=5.0cm]{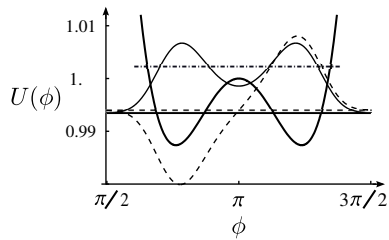}
\caption{Typical double-well configuration: three lowest energies
and two basis qubit states are shown.}\label{eigs.eps}
\end{figure}

Consider the double-well regime with $\phi_0\sim\pi$, $\alpha>1$.
Starting with the classical single-particle action
(\ref{eq:effective-action-coherent-rotating}) analytically
continued to real times, we follow standard quantization
procedure, see \cite{LeggettFrance}. While such procedure leads to
the correct result, the reader should refer to a more detailed
derivation based on the analysis of the many-body wave function
\cite{SM-JBEC}. The Lagrangian corresponding to the real-time
action is
\begin{equation}\label{eq:Lagrangian}
\frak{L} = \frac{L}{24\lambda} \dot\phi(t)^2 - E_J U(\phi(t)),
\end{equation}
where
\begin{equation}\label{eq:U_phi}
U(\phi) = \frac{1}{2\alpha}(\phi-\phi_0)^2 - \cos(\phi).
\end{equation}

The motion of the effective ``collective" particle can be formarly
quantized as follows. First, we introduce canonical momentum $P$
and perform a Legendre transformation to formulate a classical
Hamiltonian description
\begin{equation}\label{eq:U_phi}
P=\partial\mathcal{L}/\partial\dot\phi = \frac{L}{12\lambda}
\dot\phi(t),\quad\quad\quad\quad \mathcal{H} = P\dot\phi -
\mathcal{L}.
\end{equation}
We obtain
\begin{equation}\label{eq:H_phi}
\mathcal{H} = E_J\left[ \frac{1}{2\mu} P^2 + U(\phi) \right],
\end{equation}
where $\mu = \alpha\eta^2/12$, $\eta = \sqrt{\rho/m\lambda}$. The
quantization is performed introducing $P\to\hat P = -d/d\phi$, so
that $[\hat P,\phi]=1$. Finally, we arrive at the effective
Schrodinger equation describing the distribution of the collective
variable $\phi$
\begin{equation}\label{eq:H_quant}
E\psi(\phi) = E_J \hat{\mathcal{H}} \psi(\phi)
,\quad\quad\quad\quad \hat{\mathcal{H}} = \frac{1}{2\mu}
\left(-\frac{d}{d\phi}\right)^2 + U(\phi).
\end{equation}

The above quantization procedure invokes a conceptual question
\cite{amb}: The quantization is done on the quantity which is
defined by purely quantum processes, such as BEC condensation,
quantized phonon vibrations, and Josephson tunneling. Is it
appropriate to use such quantum approach to (\ref{eq:Lagrangian})
at all? A similar problem with quantization of the collective
variable takes place in the standard superconducting Josephson
ring \cite{LeggettFrance}. In the latter case (due to complexity
of the microscopic description) this question is resolved
primarily by experiment. In our case it can be done by
constructing an appropriate many-body wave function. It turns out
that the superposition of condensates, each corresponding to a
certain value of phase $\phi$, weighted by $\psi(\phi)$,
reproduces Eqs.~(\ref{eq:H_quant}) exactly in the limit of large
number of particles $1/N\ll
1/\eta\sqrt{\alpha-1}\ll\sqrt{N}/\eta$; further limitations on
$\eta$ should be enforced to maintain acceptable signal-to-noise
ratio during measurement \cite{SM-JBEC}.

The effective potential $U$ and the corresponding energy levels
are shown in Fig.~\ref{eigs.eps}. When $\phi_0\sim\pi$ the
potential acquires double well shape. The splitting energy,
$2\varepsilon$, between the two lowest energies is controlled by
the height of the barrier $\sim (\alpha-1)^{3/2}/\alpha^{3/2}$ and
effective mass $\mu$. For $\alpha-1\ll 1$ and $\eta\sim 10-100$
the energy $\varepsilon$ can be set an order of magnitude smaller
compared to the energy gap to the third level and a two state
approximation can be used. In the case of $\phi_0 = \pi$ we obtain
\begin{equation}\label{eq:Sch-Eq-qubit}
i\ket{\dot\psi} = H_0 \ket{\psi} ,\quad\quad\quad\quad H_0 =
\varepsilon \sigma_z,
\end{equation}
where state $\ket{\psi}\to (0,1)^T$ corresponds to the symmetric
ground state of (\ref{eq:H_quant}) and $\ket{\psi}\to (1,0)^T$ is
the first excited (antisymmetric) state. The splitting energy
$\varepsilon$ can be easily calculated numerically. An analytical
estimate is $\varepsilon \sim
(c/L)\exp[-2\sqrt{6}\eta(\alpha-1)^{3/2}/\alpha]$.

Finally, we should point out that the coherent description
(\ref{eq:Lagrangian},\ref{eq:H_quant},\ref{eq:Sch-Eq-qubit})
relies on the effective low-energy action
(\ref{eq:effective-action-coherent}). It was obtained assuming
$\omega L/c\ll 1$. At sufficiently large $\omega$ dissipative
corrections become important and the "coherent" description
(\ref{eq:H_quant}) is no longer valid. In the case of the
double-well configuration ($\phi_0=\pi$) $\omega$ can be estimated
as the frequency of an instanton oscillating inside the barrier
(which is potential well for the instanton). We obtain $\omega\sim
E_J\sqrt{\alpha-1}/\alpha\eta = c\sqrt{\alpha-1}/L$. Therefore the
coherent description is appropriate for small barriers when
$\alpha-1\ll 1$. On the other hand, $\alpha-1$ should not be too
small so that different $\phi$ states are still distinguishable
during measurement.

\subsection{Single-qubit operations and initialization}

Unlike in solid-state quantum computing systems, we cannot use
strong measurement procedure \cite{Nielsen} to initialize the
BEC-qubit directly. The BEC-Josephson two state system is based
upon interplay of different persistent-current states. At the same
time, measurement of BEC currents is usually destructive, as we
will see in the last section. Hence, we should resort to an
indirect measurement using a heat bath. Fortunately such
measurement is inherent to the BEC cooling process---evaporative
cooling: this process involves lowering the trap potential to
release hot ``vapor" particles; the remainder of the gas
re-thermalizes due to scattering. The preparation starts with
cooling of the trapped atomic gas. Depending on target winding
number, two strategies exist: (i) For winding numbers $<1/2$ the
pure ring potential should be used; no additional (Josephson)
barrier is initially required. (ii) For winding numbers $>1/2$, a
high impenetrable barrier should be set rotating with the
frequency $\Omega>\Omega_{1/2} = \pi^2/mL^2$.

We will focus on the case (i). Upon cooling the zero-current
($\phi\sim 0$) condensate forms. Due to the effective kinetic term
the condensate is in a superposition of different $\phi$-states
distributed in the vicinity of $\phi\sim 0$ with $\Delta\phi$
determined by the effective mass $\mu$. As soon as the Josephson
barrier is raised and set rotating by gradually increasing the
frequency of rotation to satisfy $\phi_0\approx\pi$, this
distribution will be adiabatically moved to form the ground state
of the double-well configuration, Fig.~\ref{eigs.eps}. The key is
to keep control over the angular acceleration, i.e. change of
$\Omega$ with time. In order to have the ground state fully
occupied at all times one should keep $dU(\phi)/dt \ll (E'-E_0)$,
or $\dot\phi_0\ll (E_1-E_0)$, where $E_1-E_0$ is the gap between
the first two energy levels of Eq.~(\ref{eq:H_quant}) far from the
double well configuration. We obtain $mL^2\dot\Omega/2\pi \ll
c/L$. In the vicinity of the symmetric double-well configuration
($\phi_0=\pi$), the Landau-Zener mechanism \cite{Zener} can be
used to drive the system to the antisymmetric state (similar to a
single particle double-well system).

To address single qubit operations consider the system close to
the symmetric double well configuration with $\phi_0\sim\pi$. In
the basis of the symmetric and anti-symmetric states the
Hamiltonian takes the form
\begin{equation}\label{eq:Sch-Eq-qubit-full}
H = \varepsilon \sigma_z +
\frac{\phi_0\av{\phi}_{01}}{\alpha}\sigma_x,
\end{equation}
where $\av{\phi}_{01}$ is the off-diagonal matrix element of
$\phi$ in the same basis. Both phase and flip \cite{Nielsen}
operation can be carried out by tuning the parameters
$\varepsilon$ and $\phi_0$ (i.e. the Josephson tunneling and
frequency of rotation of the barrier).

\section{Multi-qubit dynamics}\label{sec:MQ}\label{sec:Measure}

It is well known that superconducting flux qubits can be easily
coupled via the magnetic field \cite{squid} generated by
supercurrent in each qubit. However, all the currents in the
cold-atom-based qubit are flows of neutral particle. Hence, they
do not interact with electromagnetic fields. The only interaction
present in the system is contact scattering. Therefore
BEC-Josephson rings must come into contact with one-another or
with some common (coherent) BEC media. Below we discuss the
simplest stack arrangement based on this idea. While more advanced
schemes are possible, we will focus on demonstrating the basic
principles necessary to construct such an interaction in the
simplest possible model. Multi-plane geometry, however, is not
easy to implement with more than a few planes---investigation of
more complex (in-plane) coupling schemes might be necessary to
address the issue of scalability.
\begin{figure}
\includegraphics[width=3.0cm]{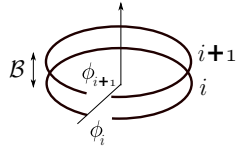}
\caption{Stack arrangement: BEC-Josephson rings are placed one on
top of the other, so that small tunneling is present between the
rings. The energy of the system is minimized when the rings are in
phase.}\label{rings.eps}
\end{figure}

\subsection{Interaction between a few BEC rings}

Consider a stack of $M$ similar BEC-Josephson rings placed one on
top of the other in $z$ direction (perpendicular to the 2D
confinement planes) with the common axis of rotation and the same
rotation frequency. The separation between the rings is such that
weak tunneling is present between the equivalent parts of the
adjacent rings, see Fig.~\ref{rings.eps}. The Euclidian action
corresponding to such configuration can be cast in the form
\begin{equation}\label{eq:Sn0-def}
\S_M=\S^M_0+\S^M_\J+\S^M_\B.
\end{equation}
The three components are: (i) The effective 1D action describing
the bulk of each BEC ring
\begin{equation}\label{eq:Sn0-def}
\S^M_0=\sum_{i=1}^M\int_0^\beta d\tau\int_0^Ldx
\psi_i^*(x,\tau)\left[
\partial_\tau -\frac{\nabla^2}{2m} +
\frac{\lambda}{2}\psi_i^*(x,\tau)\psi_i(x,\tau)\right]\psi_i(x,\tau),
\end{equation}
(ii) The Josephson tunneling contribution for each ring in a stack
\begin{equation}\label{eq:SnJ-def}
\S^M_\J = \sum_{i=1}^M\int_0^\beta
d\tau\J\left[\psi_i^*(0,\tau)\psi_i(L,\tau) +
\psi_i^*(L,\tau)\psi_i(0,\tau)\right],
\end{equation}
(iii) The tunneling between adjacent rings in the stack
\begin{equation}\label{eq:SnB-def}
\S^M_\B = \sum_{i=1}^{M-1}\int_0^\beta d\tau\int_0^Ldx
\frac{\B}{L}\left[\psi_i^*(x,\tau)\psi_{i+1}(x,\tau) +
\psi_{i+1}^*(x,\tau)\psi_i(x,\tau)\right].
\end{equation}
Here $\B$ is the tunneling amplitude between adjacent rings. As
before we assume that both $E_J=\J\rho$ and $E_B=\B\rho$
(Josephson energy due to the tunneling between the rings) are
small compared to the quantized phonon excitation energies,
$\min(ck)=2\pi c/L$, in the bulk. We will also assume that all the
rings in the stack are arranged such that the Josephson barriers
are on top of each other. In such system the phases at the same
point, $x$, along the rings do not change significantly between
different rings $\varphi_i(x,\tau)\sim
\varphi_{i+1}(x,\tau)$---the variation of $\phi_i$ for each qubit
state is relatively small ($\phi_i\sim\pi$) and
$\varphi_i(x,\tau)\approx \phi_i x/L$ with small fluctuations
around this value. We can use the hydrodynamic parametrization as
before $\psi_i(x,\tau)=\sqrt{\rho + \delta\rho_i(x,\tau)}
e^{i\varphi_i(x,\tau)}$ and, to the leading order in fluctuations,
obtain
\begin{equation}\label{eq:Sn-expanded}
\S_n =\sum_i\S^i_\ind{MF} + \sum_i\int dxd\tau  \left[
i\delta\rho_i\dot\varphi_i + \frac{\rho}{2m}(\nabla\varphi_i)^2 +
\frac{\lambda}{2}\delta\rho_i^2 -
\frac{E_B}{L}\cos(\varphi_i-\varphi_{i+1}) \right] - \sum_i\int
d\tau E_J\cos\phi_i.
\end{equation}
Here (and in the following) we suppress the arguments of the
fields and the limits of the integration to shorten notation. We
also expand $\S_n$ to the second order in $\varphi_i(x,\tau) -
\varphi_{i+1}(x,\tau)$. Higher-order corrections are irrelevant
since $E_B$ is a small quantity itself. Integrating out the
density fluctuations for each ring as before, we obtain
\begin{equation}\label{eq:Sn-phonons}
\S_n =\sum_i\S^i_\ind{MF} +
\frac{1}{L}\sum_{i,k}\int\frac{d\omega}{2\pi} \left[
\frac{\omega^2+c^2k^2}{2\lambda}|\varphi_i|^2  +
\frac{E_B}{L}|\varphi_i-\varphi_{i+1}|^2 \right] - \sum_i\int
d\tau E_J\cos\phi_i.
\end{equation}
Further integration is more involved but follows the same steps as
in the case of the single BEC-Josephson ring: (i) the phases
$\phi_i$ across the junction are introduced inserting $\delta [
\phi_i(\tau) - \{\varphi_i(0,\tau)-\varphi_i(L,\tau)\}]$; (ii) the
$\delta$-functions are expanded using the identity
$\delta(a_i)=(2\pi)^{-1}\int d\Lambda_i \exp{(i\Lambda_i a_i)}$;
(iii) finally, the fields $\varphi_i$ and then $\Lambda_i$ are
integrated out (Gaussian integrals) in the same manner as before.
We should note that the Gaussian integrals over $\varphi_i$ are
not diagonal in the index that counts different rings, hence, a
diagonalization has to be done to perform the integration.
Fortunately we do not have to compute these integrals since they
give only a constant shift to the free energy (action) of the
system which is not important here. As soon as phonon and
auxiliary fields are integrated out the second sum of action
(\ref{eq:Sn-phonons}) turns into
\begin{equation}\label{eq:K1K2-fragment}
\sum_i\int \frac{d\omega}{2\pi} \left[
K_1(\omega)|\phi_i(\omega)|^2 + E_B
K_2(\omega)|\phi_i(\omega)-\phi_{i+1}(\omega)|^2  \right],
\end{equation}
where
\begin{equation}\label{eq:K1}
K_1(\omega)^{-1} = \frac{1}{L}\sum_q
\frac{(1-e^{-iqL})^2}{(\omega^2+c^2q^2)/2\lambda} =
\frac{4\lambda}{c} \frac{\tanh\frac{L\omega}{2c}}{\omega} %
\end{equation}
and
\begin{equation}\label{eq:K2}
K_2(\omega)^{-1} = \frac{K_1(\omega)^2}{L}  \frac{1}{L}\sum_q
\left[ \frac{1-e^{-iqL}}{(\omega^2+c^2q^2)/2\lambda}\right]^2
\stackrel{\omega\to 0}{\longrightarrow} \frac{1}{12}.
\end{equation}
Following Sec. \ref{sec:BECJ} we keep only the leading frequency
terms. Hence, only the zeroth order in $\omega L/c$ should be
retained in $K_2(\omega)$. After some algebra we obtain the
effective low-energy action describing the entire stack
\begin{equation}\label{eq:Sn-eff-final}
\S^M_\ind{eff} = \sum^M_{i=1}\int d\tau \left[
\frac{L}{24\lambda}\dot\phi^2 + \frac{\rho}{2mL}\phi_i^2 + \rho
v\phi_i + E_J\cos\phi_i +
\frac{1}{12}E_B(\phi_i-\phi_{i+1})^2(1-\delta_{i,M}) \right].
\end{equation}
The form of this action is not surprising. Indeed, by allowing the
tunneling between the rings we increase the energy of the entire
system if those rings are not in phase. This is manifested (to the
leading order in the phase difference) by the appearance of the
last quadratic term in the action, see
Eq.~(\ref{eq:Sn-eff-final}).

\subsection{Interacting qubits}

We have demonstrated that a set of $M$ BEC-Josephson qubits can
interact with the total effective energy as a function of the
phase-slip phase difference given by
\begin{equation}\label{eq:Un_phi_nm}
U(\phi_1,\phi_2,...) = \sum_{i=1}^M\left[
\frac{1}{2\alpha}(\phi_i-\phi_{0})^2 - \cos(\phi_i) \right] +
\frac{\B}{\J}\sum_{i=1}^{M-1}(\phi_i-\phi_{i+1})^2.
\end{equation}
However, it is still necessary to prove that potential
(\ref{eq:Un_phi_nm}) is sufficient to implement a universal set of
gates \cite{Nielsen}. Below we demonstrate that qubit-qubit
coupling given by (\ref{eq:Un_phi_nm}) together with a single
qubit rotation discussed earlier can be used to implement any
two-qubit gate.

Consider a stack of only two BEC-Josephson rings. In the basis of
symmetric/antisymmetric states the Hamiltonian becomes
\begin{equation}\label{eq:Hn-qubits-all}
H = H_1 + H_2 + \frac{2\B\av{\phi}_{01}^2}{\J}
\sigma^1_x\sigma^2_x,
\end{equation}
\begin{equation}\label{eq:Hn-qubits-i}
H_i = \varepsilon\sigma^i_z + \left(\frac{\phi_0-\pi}{\alpha} +
\frac{2\pi\B}{\J} \right)\av{\phi}_{01}\sigma^i_x.
\end{equation}
We tune $\varepsilon\to 0$ and $\phi_0\to\pi - 2\pi\alpha\B/\J$.
In this case the evolution operator (in real time) is
\begin{equation}\label{eq:U-2qubits}
U(t) = U^{1\dag}_{x\sigma}U^{2\dag}_{x\sigma}\exp\left[ -i \frac{2
t\B\av{\phi}_{01}^2}{\J}\, \mathbf{\sigma}^1\cdot\mathbf{\sigma}^2
\right]U^1_{x\sigma}U^2_{x\sigma}.
\end{equation}
Apart from single qubit rotation $U_{x\sigma}$: $\sigma_x\to\sigma
=\{\sigma_x,\sigma_y,\sigma_z\}$ this is a natural representation
\cite{SWAP} of a (SWAP)$^\xi$ gate with $\xi = {2
t\B\av{\phi}_{01}^2}/{\J}$. This gate, in combination with
single-qubit rotations, is sufficient to implement any two-qubit
gate \cite{SWAP}. For instance, one CNOT gate can be realized with
the help of two (SWAP)$^{1/2}$ and single-qubit gates \cite{Loss}.
\begin{figure}
\includegraphics[width=10.0cm]{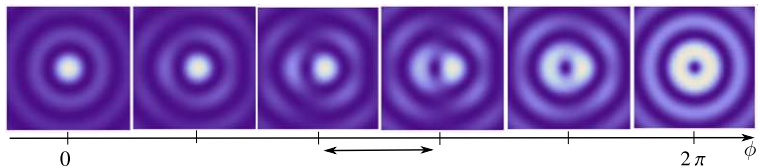}
\caption{Calculated Time-of-Flight (TOF) measurement outcomes.
During the TOF measurement BEC-Josephson qubit wave function
should collapse to a certain $\phi$ state. The density
distribution of that state can be captured by absorbtion imaging
during the TOF expansion. The two-way arrow indicates typical
range of $\phi$ expected for measurable two-state BEC-Josephson
system.}\label{measure-all.eps}
\end{figure}
\begin{figure}
\includegraphics[width=5.0cm]{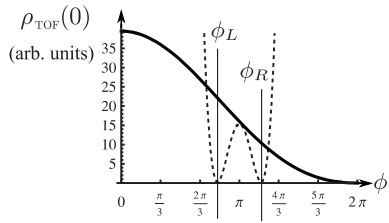}
\caption{Time-of-Flight density (calculated) at the center as a
function of the phase-slip $\phi$ of the ring system before the
expansion. A typical effective potential corresponding to
BEC-Josephson qubit configuration is shown for comparison
($\phi_{L/R}$ marks the average position of states in the
effective two state system).}\label{TOF-DW.eps}
\end{figure}

\section{Measurement}\label{sec:Measure}

The only reliable mechanism to measure a cold atom system is via
observation of its density. Moreover the atomic cloud has to
undergo significant expansion before images with sufficient
spacial resolution can be obtained. The latter is usually
accomplished by Time-of-Flight (TOF) experiments
\cite{LeggettBEC}: the trapping potential is turned off and the
density of the atomic cloud is measured after some time of free
expansion. As soon as the confinement potential is off, the system
is just a collection of atoms that will fly away from the position
of the trap, each with the momentum supplied by the confinement
and motion within the trap. Due to the tight confinement, the
released atoms acquire substantial velocities which triggers fast
expansion and decrease of density. The scattering between
particles can be neglected, since it usually leads to effects on a
much longer time scale. The wave function of the free-expanding
BEC is
\begin{equation}\label{eq:PsiTOF}
\Psi(\mathbf{r}_1,\mathbf{r}_2,...,t) = \int
\frac{d\mathbf{p}_1}{(2\pi)^3}\frac{d\mathbf{p}_2}{(2\pi)^3}...
\Psi(\mathbf{p}_1,\mathbf{p}_2,...)e^{i\mathbf{r}\mathbf{p}-i(p^2/2m)t},
\end{equation}
At sufficiently large times only terms with $\mathbf{r}\sim
\mathbf{p}t/2m$ contribute to the integral. As the result, TOF
measurement provides an image of the density in momentum space.
This is ideal to capture the distribution over different current
states present in (\ref{eq:H_quant}). Consider different moments
of density in the momentum space $m_n=\av{
\hat\rho(\mathbf{q}_1)...\hat\rho(\mathbf{q}_n) }$, where
$\hat\rho(\mathbf{q}) = \int
d\mathbf{r}d\mathbf{r}'e^{i\mathbf{q}(\mathbf{r}-\mathbf{r}')}
\hat\psi^\dag(\mathbf{r}')\hat\psi(\mathbf{r})$,
\begin{equation}\label{eq:m_n-def}
m_n = \int \D\psi^*\D\psi\int
d\mathbf{r}_1d\mathbf{r}_1'...d\mathbf{r}_nd\mathbf{r}_n'
e^{i\mathbf{q}_1(\mathbf{r}_1-\mathbf{r}'_1)}...
e^{i\mathbf{q}_n(\mathbf{r}_n-\mathbf{r}'_n)}
\psi^*(\mathbf{r}_1',0)\psi(\mathbf{r}_1,0)...
\psi^*(\mathbf{r}_n',0)\psi(\mathbf{r}_n,0) e^{-\S}.
\end{equation}
Here $\S$ is given by Eqs.~(\ref{eq:S0-def}) and
(\ref{eq:SJ-def}). The expansion in terms of small fluctuations
$\psi(\mathbf{r},\tau) = \chi_{\phi(\tau)}(\mathbf{r}) +
\delta\psi(\mathbf{r},\tau)$ yields
\begin{equation}\label{eq:m_n}
m_n = \int d\phi P(\phi)
|\chi_\phi(\mathbf{q}_1)|^2...|\chi_\phi(\mathbf{q}_n)|^2 +
g^2[f(\mathbf{q}_1,\mathbf{q}_2)+f(\mathbf{q}_2,\mathbf{q}_3)+...]+\O(g^4),
\end{equation}
where the (path) integrals over $\phi$ with $\tau\neq 0$ has been
lumped into $P(\phi)$. The function $f(\mathbf{q},\mathbf{q}')$ is
a combination of two particle Green's functions describing
scattering processes in and out of the condensate, see \cite{AGD}.
These two-particle correlations are suppressed \cite{LeggettBEC}
by the gas parameter $\sim g$, which limits the resolution of the
measurement. For the range of parameters appropriate for the
macroscopic two state BEC-Josephson system the gas parameter is
small. Apart from the $g^2$ correction, moments~(\ref{eq:m_n})
define a stochastic process: $P(\phi)$ plays the role of the
probability to find the system at state $\phi$ as the result of
the measurements. Hence, by measuring the density distribution in
momentum space (via TOF) we measure the value of $\phi$. At each
experimental run a different density distribution is expected with
the probability determined by $|\psi(\phi)|^2$.

The possible TOF measurement outcomes are given in
Fig~\ref{measure-all.eps}, where we plot the momentum-space
density distributions $|\chi_\phi(\mathbf{q})|^2$ for several
values of $\phi$ from 0 to $2\pi$. Different $\phi$ states are
clearly distinguishable. For a more quantitative judgment we
propose to measure the TOF density at the center of the trap. The
central TOF density, $|\chi_\phi(0)|^2 = |\int d\mathbf{r}
\chi_\phi(\mathbf{r}) |^2$, depends on the phase
$\varphi(\mathbf{r})$ in a straightforward way. In the bulk of the
ring this phase is linear (in this case kinetic energy is
minimized) $\varphi(\mathbf{r}) = \phi x/L$ and, hence,
$\chi_\phi(\mathbf{r})\sim e^{i\phi x/L}$. As the result we obtain
\begin{equation}\label{eq:m_n}
\rho_{TOF}(0) =
\rho^{\phi=0}_{TOF}(0)\frac{\sin^2(\phi/2)}{(\phi/2)^2}.
\end{equation}
For the variation of $\phi$ relevant to the BEC-Josephson qubit
system, see Fig~\ref{TOF-DW.eps}, this is an approximately linear
function $4/\pi^2 - 8(\phi-\pi)/\pi^2$.

In this review we have considered only the very basic properties
of the cold atom BEC-Josephson rings relevant to quantum
information. A detailed investigation is necessary to assess noise
effects as well as scalability. A simple proof-of-principle
experiment with a single BEC-Josephson macroscopic two-state
system should also be helpful in determining the direction for
further theoretical development.

{\acknowledgements We thank M. G. Boshier, I. Martin, V. Privman
and E. Timmermans for valuable discussions and comments. DS
acknowledges stimulating conversations with R. Kalas. The work is
supported by the US DOE. }


\end{document}